\documentclass[12pt]{spieman}  
\usepackage{amsmath,amsfonts,amssymb}
\usepackage{graphicx}
\usepackage{setspace}
\usepackage{tocloft}
\usepackage{subcaption}

\title{Expected performances of a Laue lens made with bent crystals}

\author[a,*]{E.~Virgilli}
\author[b]{V.~Valsan}
\author[a,c]{F.~Frontera}
\author[c]{E.~Caroli}
\author[d]{V.~Liccardo}
\author[c]{J.B.~Stephen}

\affil[a]{Department of Physics, University of Ferrara, Via Saragat 1/c, 44122 Ferrara, Italy}
\affil[b]{Indian Institute of Astrophysics, Koramangala, Bangalore - 560034, India}
\affil[c]{IASF-Bologna, INAF, Via Gobetti 101, 40129 Bologna, Italy}
\affil[d]{ITA-Instituto Tecnol\'ogico de Aeron\'autica, S\~ao Jos\'e dos Campos, Brasil}

\cftpagenumbersoff{figure}
\cftpagenumbersoff{table}

\newcommand{\psf}{{\sc psf}}
\newcommand{\hpd}{{\sc hpd}}
\newcommand{\fwhm}{{\sc fwhm}}

\newcommand{\laue}{{\sc laue}}

\newcommand{\larix}{{\sc larix}}
\newcommand{\llll}{{\sc lll}}

\newcommand{\qm}{{\sc qm}}

\begin{document} 
\maketitle

\begin{abstract}
In the context of the \laue\ project devoted to build a Laue lens prototype for focusing celestial hard X-/soft gamma-rays, a Laue lens made of  
bent crystal tiles, with 20 m focal length, is simulated.
The focusing energy passband is assumed to be 90--600 keV.
The distortion of the image produced by the lens on the focal plane, due to effects of 
crystal tile misalignment and radial distortion of the crystal curvature, is investigated.
The corresponding effective area of the lens, its point spread function and sensitivity are calculated and compared with those exhibited by a nominal Laue lens with no misalignment and/or distortion. Such analysis is crucial to estimate the optical properties of a real lens, in which the investigated shortcomings could be present.
\end{abstract}


{\noindent \footnotesize\textbf{*}Corresponding author:  \linkable{virgilli@fe.infn.it}~~~{\bf larixfacility.unife.it}}

\maketitle

\section{Introduction}
\label{intro}

Motivated by the astrophysical importance  of extending the focusing band  up to at least 600 keV~\cite{Frontera10}, a project named \laue\ 
was approved and supported by the Italian Space Agency, with the goal of finding a 
well-grounded technology for building Laue lenses~\cite{Frontera12} with a broad energy passband.
After preliminary Monte Carlo simulations, a Laue lens prototype made of  bent crystals, with an energy passband from 90 to 300 keV and a focal length of 20~m, is being developed in the \larix\ facility~\cite{larix}  of the Physics and Earth Sciences Department of the University of Ferrara. 
Along with the experimental development activity which is ongoing in our Institutes, 
we are starting the feasibility study of an instrument concept, {\bf ASTENA} ({\em 
Advanced Surveyor of Transient Events and Nuclear Astrophysics}), supported by the 
European  project {\bf AHEAD} ({\em Integrated Activities for the High Energies 
Astrophysics Domain}\footnote[3]{\href{http://ahead.iaps.inaf.it/}{http://ahead.iaps.inaf.it/}}. 
The instrument includes a wide field monitor/spectrometer (1~keV -- 20~MeV ) 
and an optimized narrow field  telescope (NFT), made of a 3~m diameter broad-band Laue lens 
(50--700~keV) with a 20~m focal length, coupled with a focal plane position sensitive detector 
with 3D spatial resolution.

For the first time, bent crystals of Ge(111) and GaAs(220) in transmission (or Laue) configuration are being used.
Their curvature $c$ (also called {\em external curvature} or {\em primary curvature}) with radius $r=1/c$, is obtained with 
mechanical processes of lapping~\cite{buffagni11} in the case of GaAs(220), and grooving~\cite{guidi13} in the case of Ge(111). 
Crystals with bent diffractive planes, when compared to their flat mosaic counterparts, have 
very interesting properties. On one side, bent crystals have the valuable capability of concentrating the parallel polychromatic 
beam into a focal spot which is smaller than the single crystal cross section itself, while flat crystals produce a diffracted image having at least the same 
size of the crystal tile. In addition, bent crystals are expected to show a higher efficiency than the maximum value (50\%) expected for flat crystals made 
of the same material~\cite{malgrange2002}.
Experimental campaigns performed on bent crystals of Ge(111)~\cite{liccardo14} have confirmed the expectations.

For some crystallographic orientations (e.g. for the (111) planes of perfect Germanium crystals),  
the external curvature induces a $secondary~curvature$ of the internal diffractive planes and we
call $r_s$ its correspondent curvature radius. 
The secondary curvature enlarges locally the crystal energy bandwidth with respect to that of a perfect crystal. 
This widening of the passband is somehow similar to the mosaicity $\omega_m$ of flat mosaic crystals. Indeed it is known as  
{\it quasi-mosaicity}\cite{sumbaev57, Ivanov05} and the secondary curvature 
is also called {\it quasi-mosaic} (\qm) curvature.
The \qm\ is a consequence of the crystal anisotropy~\cite{lekhnitskii56, sumbaev68} and the relation between primary and \qm\ curvature can be estimated through the 
linear theory of elasticity. For crystals made of Ge(111), it has been empirically estimated to be 
$r_s = -2.39~r$~\cite{Bellucci13}.

The technology to produce bent crystals with the proper primary curvature radius is still in a R\&D phase to identify materials and bending procedures for achieving the desired curvature. 
To date, for the \laue\ project, crystal tiles of Ge(111) and GaAs(220) have been bent with curvature radii within 5-10$\%$ of the desired value (40 m).
Therefore it is important to evaluate the performances of a Laue lens
made of crystal tiles whose curvature radii are spread in a specific range centered at 
the nominal curvature radius. We refer to this shortcoming as the {\em radial distortion}.

Another crucial aspect to be tackled is the positioning accuracy of each crystal tile on the lens frame. 
Setting each crystal at the proper position and orientation to diffract 
the photons at the lens focal point requires a dramatically accurate process, 
and possible deviations of the crystals from their proper position ({\em angular misalignment}) have to be 
considered.

In this paper, using a Monte Carlo code, we derive the expected performance of a lens, in
which, in addition to the nominal case of properly bent tiles oriented as required for perfect
focusing, we consider the case of crystal tiles with radial distortion and/or angular misalignments.
The knowledge of how these shortcomings reduce the lens performances allows to develop
a strategy to minimize their effects. 

The paper is organized as follows. In Sect.~\ref{s2} it is given a geometric description of the spatial distribution 
of the diffracted photons on the focal plane detector  for both the effects of  crystal tile misalignment
and radial distortion. In Sect.~\ref{s3} is described the developed software to 
simulate the overall behavior of a Laue lens made of bent crystals, each one with a possible radial uncertainty 
and/or angular misalignment. In Sect.~\ref{s4} the simulated lens, made of
bent Ge (111) crystals, is described while in Sect.~\ref{s5} the simulation results are discussed.
For each configuration 
we have derived  the lens performances and the results have been compared with these obtained with a 
nominal lens (i.e. made with properly bent tiles oriented as required for a perfect focusing).
Using the Monte Carlo method, the spread of the photons in the focal plane has been found in excellent 
agreement with the geometric description of the shortcomings.
Finally, in Sect.~\ref{s6} we have drawn our conclusions and given some prospects 
for future activities related to the employ of Laue lenses for X-/Gamma-ray observations.

\section{Geometric description of the crystal-misalignement and radial-distortion effects}
\label{s2}

In this section a geometric description of the effects on the spatial distribution 
of the diffracted photons is given for both the effects of mounting accuracy of a 
crystal tile on the lens frame, and the effect of a not proper primary curvature radius. It is worth noting
that the following treatise explains the spatial distribution of the photons diffracted by a single crystal 
affected by these sources of uncertainties. The overall effect in
a Laue lens is given by the superposition of all the crystals contributions, and this is rigorously
performed with the Monte Carlo method presented in Sect.~\ref{s3}.

\subsection{Crystal tile misalignment}
\label{effect_misalignment}

\begin{figure}[!t]
  \centering
  \includegraphics[scale=0.55]{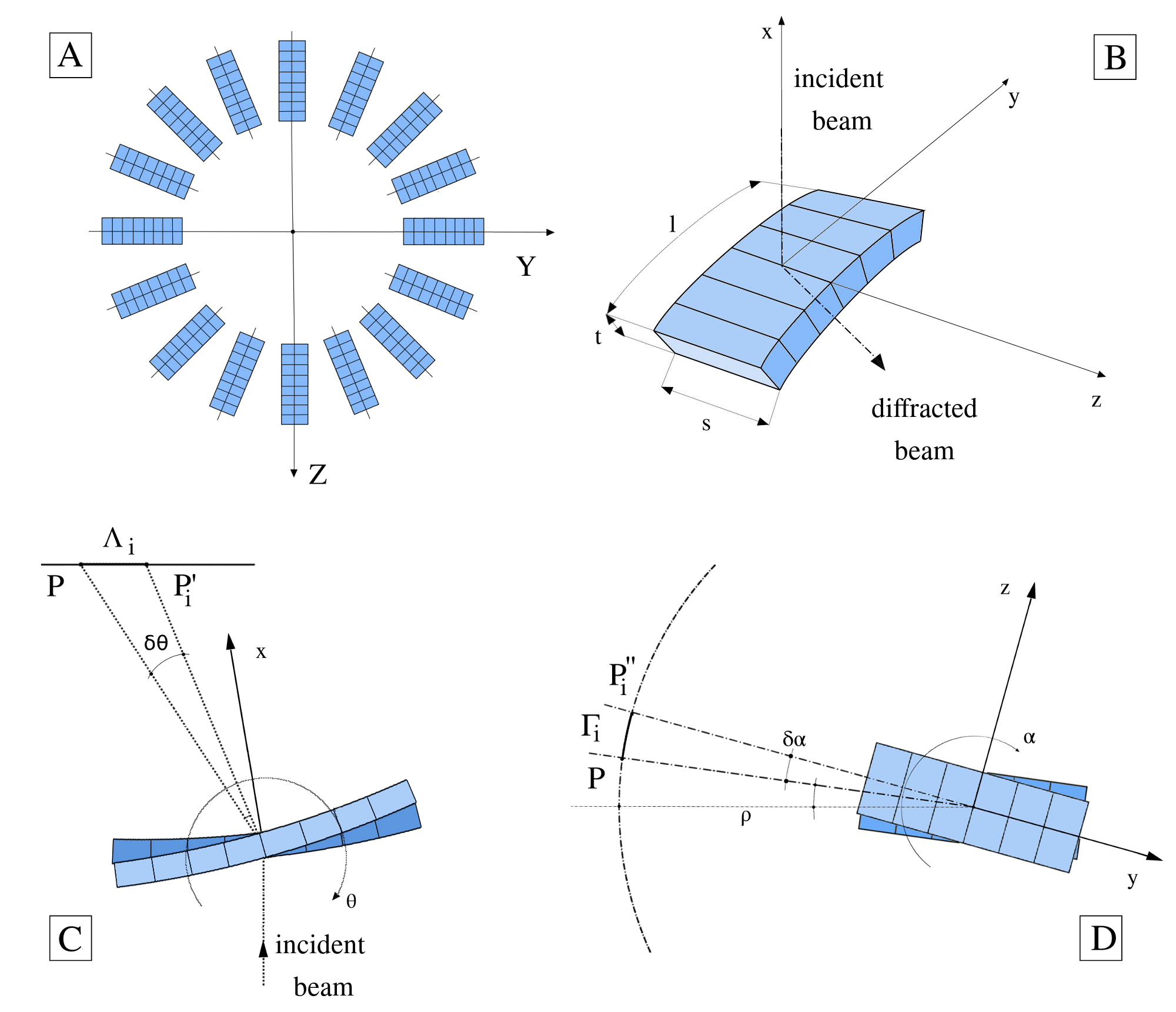}
  \caption{$A$: Sketch showing a fraction of Laue lens (a single ring of crystals), the 
  orientation of the crystals and the diffracting planes. The Y and Z axes represent the laboratory reference
  frame. $B$: The crystal reference axes used in this paper. The dimensions of each crystal 
  (dimension $l$ along the primary curvature direction, dimension $s$ along the non-focusing 
  direction $z$, tile thickness $t$) are also shown. $C$: Side view of a crystal. Under 
  a tilt from $\theta$ to  $\theta + \delta\theta$ the diffracted beam from the  i$^{th}$ crystals shifts by an amount 
  $\Lambda_i$ from  the nominal lens focus $P$  to $P_i'$. $D$: Front view of a crystal. When the i$^{th}$ crystal undergoes a tilt 
  around the $x$ axis from $\alpha$ to $\alpha + \delta \alpha$,  the diffracted beam centroid 
  shifts along the indicated arc from the nominal lens position $P$ (center of the lens, see panel A) to $P_i''$ by an amount $\Gamma_i$ proportional to the distance 
  of the i$^{th}$ crystal from the lens axis ($\rho_i$).}
  \label{fig:misalignment}
\end{figure}

In a Laue lens each crystal tile has to be properly oriented
so that the diffracted photons by each crystal arrive in the lens focus.
If the crystals are misaligned from their nominal orientation, the corresponding diffracted photons
will be shifted with respect to the lens focus. With reference to Fig.~\ref{fig:misalignment}, 
the orientation of each crystal is given by the three angles $\alpha$, $\phi$, and $\theta$ 
around the main crystal axes $x$, $y$ and $z$, respectively (see panel B).

On first approximation, a rotation $\phi$ around the $y$ axis does not affect 
the position of the diffracted photons. This assumption is true if the diffractive planes are perpendicular to the $y$ axis, as assumed. Therefore 
a misorientation over the $\phi$ angle will be considered negligible.
Let consider the $i^{th}$ crystal. When the crystal is correctly oriented with angles
$\theta_i$ and $\alpha_i$, the centroid of its diffracted beam will be {\bf P}. Any variation $\delta\theta$ and $\delta\alpha$ of these angles shifts the diffracted beam 
to a different position. 

For a misalignment $\delta\theta$, the new position is $P_i'$ with a linear deviation $\Lambda$ given by:

\begin{align}
\Lambda_i \sim f  ~ \delta\theta
\end{align}

 where $f$ is the focal length of the Laue lens. 

Similarly, a change of the azimuthal angle from 
$\alpha_i$ to $\alpha_i$+$\delta\alpha$ will results in a new position of the diffracted beam, with a shift $\Gamma_i$
along the arc shown in Fig.~\ref{fig:misalignment} (panel D),  given by:

\begin{align} 
\Gamma_i \sim \rho_i ~\delta\alpha
\end{align}

where $\rho_i$ is the distance of the $i^{th}$ crystal from the lens axis. 
Both the angle $\delta\theta$ and $\delta\alpha$ have been 
incorporated in the code. It is worth noting that, for astrophysical 
applications, the focal length is 10-20 m and the rings 
have a radius $\rho_i < 2$~m, hence the misalignment effect 
caused by the tilt around the $z$ axis is more pronounced
than that linked to the tilt around the $x$ axis.

\subsection{Radial distortion}
\label{effect_radial}

Independently of the process used to cause the external curvature (and consequently the secondary one), the crystal curvature 
radius can deviate from the nominal radius. Therefore, a systematic study of
the Point Spread Function (\psf) dependence on the deviation of the crystal primary curvature from the 
nominal one is very important. A qualitative effect of this radial 
distortion is illustrated in  Fig.~\ref{fig:radialDisto}. As can be seen, in this case, the distorted crystals are 
oriented in such a way that the diffracted photons are 
focused in the best way possible.    

Let consider a bent crystal of Ge(111) with the sizes shown in Fig.~\ref{fig:misalignment} (panel B).
Let call W$_{a, b}$ the Full Width at 
Half Maximum (\fwhm) of its \psf\ where
the first index $a$ indicates the crystal curvature radius and $b$ indicates 
the distance from the crystal at which the \fwhm\ is measured.
By keeping in mind that, for geometric reasons, the focal length is a half of the primary
curvature radius of the crystal ($f = r/2$), by using the relation $r_s = -2.39~r$ seen in Sect.~\ref{intro}, it can be easily shown that the \fwhm\ of the \psf\ of a single crystal tile is given by:

\begin{equation}
W_{r,r/2} = 2 ~ f ~ \Omega = 2 ~ \frac{r}{2}~ \frac{t}{r_s} \sim \frac{t}{2.39} 
\label{width2.39}
\end{equation}

where $\Omega$ is the total bending angle of the planes corresponding to the quasi--mosaicity ($\Omega = \omega_m$) and $t$ is the crystal 
thickness. Thus, on first approximation, from the above relation the \fwhm\ depends only on the crystal thickness and is valid for each radius. 
Then, for a crystal with radius 
r$^*$ = r + dr it results that W$_{r^*,r^*/2}$  = W$_{r,r/2}$.
In particular, for the case of r$^*$ = r + dr (and similarly for the case r$^{**}$ = r - dr), from simple geometric considerations 
(see Fig.~\ref{fig:radialDisto}) we get:

\begin{figure}[!t]
  \centering
    \includegraphics[scale=0.24]{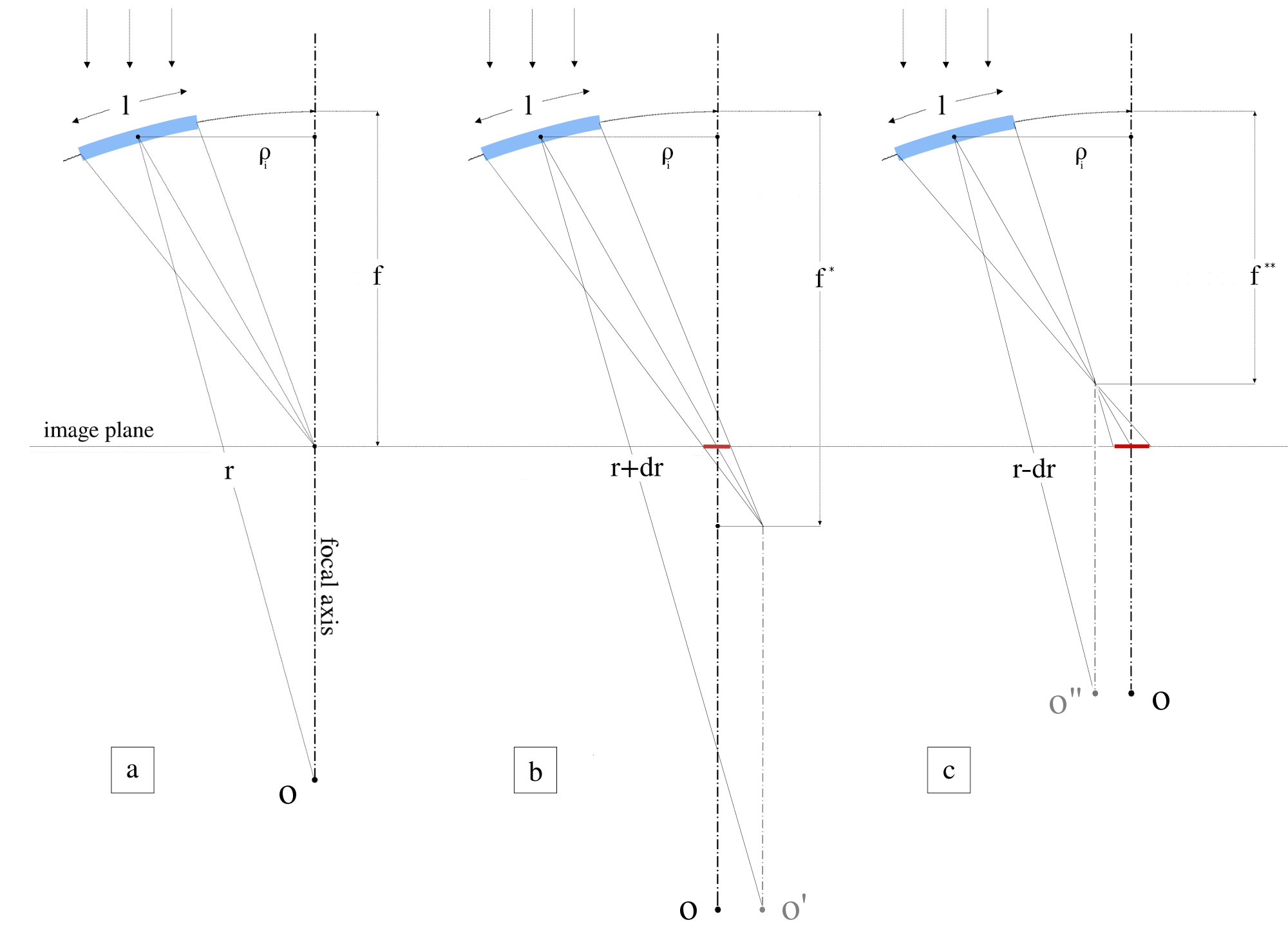}
  \caption{A qualitative representation of the focusing effect in the case of a crystal with the proper curvature
  radius $r$ ($a$) compared with the cases of crystal with primary curvature radius $r$+$dr$ ($b$) and  $r$-$dr$ ($c$).}
  \label{fig:radialDisto}
\end{figure}

\begin{equation}
W_{r^*,r/2} = W_{r,r/2} + \alpha^*~|f^* - f| \sim W_{r,r/2} ~ + ~\alpha^*~ \frac{dr}{2}
\end{equation}

where $f^*=r^*/2$ and the angle $\alpha^*$ is given by:

\begin{align}
\alpha^* = \frac{l ~~ cos~\theta_B}{f^*} \sim \frac{l}{f+\frac{dr}{2}}
\end{align}

The last approximation ($cos~\theta_B \sim 1$) is justified by the fact that, for the simulated Laue lens with 20~m focal length, even the maximum Bragg angle, corresponding to the minimum energy (90 keV), is very small ($\sim$1.2$^o$). Thus it results:

\begin{align}
W_{r^*,r/2} \sim \frac{t}{2.39} + l ~ \frac{dr}{r + dr}
\label{W}
\end{align}

According to the Eq.~\ref{W},  when no radial distortion is present with respect to the nominal 
radius, the second term disappears and the only contribution to the \fwhm\ is given by the \qm\ effect. 
On the contrary, if a radial distortion is present, the second part dominates the \fwhm\ estimation.

\section{Description of the ray-tracing code}
\label{s3}

\begin{figure}[!t]
  \centering
  \includegraphics[scale=0.9]{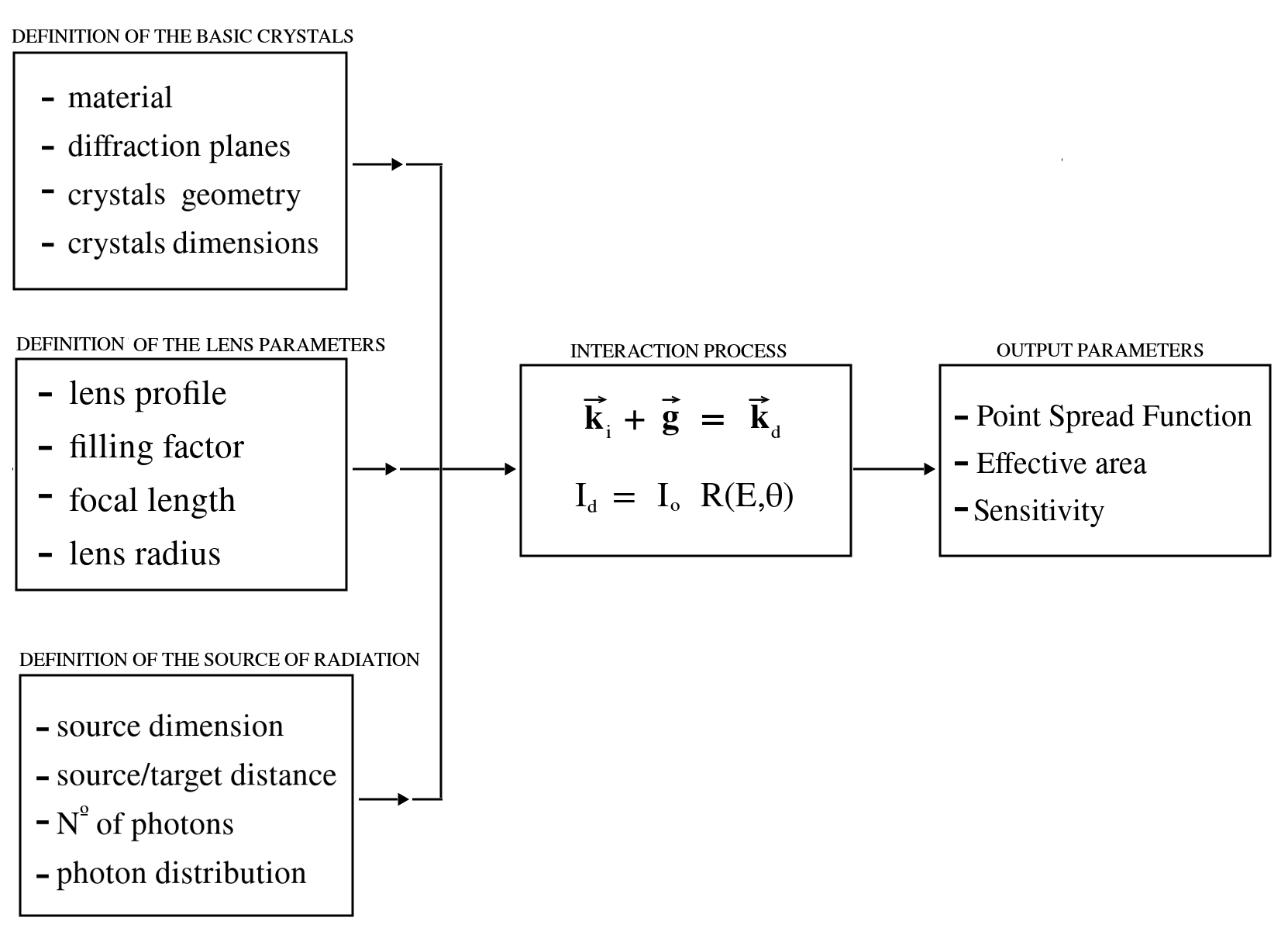}
  \caption{Diagram showing the working principles of the Laue Lens Library (\llll).}
  \label{fig:diagram}
\end{figure}

The code used for the Monte Carlo simulations is developed in Python.  It describes the lens geometry and the 
diffraction process which concentrates the incident photons on the lens crystals  towards the lens focal point.
The lens is made of a spherical cup filled with crystal tiles distributed in concentric rings
around the lens axis, as described, e.g., in Ref.~\citenum{Frontera10}.
 The software consists of a number of functions 
each responsible of a given task (photon production, crystal definition, 
lens geometry, physics of the processes, data acquisition). With reference to the block diagram 
shown in Fig.~\ref{fig:diagram}, the user interacts with the code providing the 
parameters required for the definition of the lens properties. 

The crystal tiles can be made of a single or more materials, while the diffraction planes of each crystal material are defined through the  Miller indices. 
The user must provide the crystal dimensions ($l$ along the crystal curvature, $s$ along the normal (non-focusing) direction and $t$ the crystal thickness), while 
the tile spatial position (${x_i}$, ${y_i}$, ${z_i}$) and the orientation angles ($\alpha_i$, 
${\theta}_i$, $\phi_i$) are independently calculated by the software once the Laue lens radial extension and focal length are provided.
The crystals can be either flat or bent and, in the latter case, the curvature radius
of each tile must be provided. Each crystal can be correctly oriented at its nominal 
position or misaligned within a given range 
with respect to the nominal orientation. Moreover, the crystal curvature 
radius can be either set at the nominal value or distributed over a range of curvature radii centered 
at the nominal radius and following a uniform or a Gaussian distribution.
Depending on the lens energy passband and on the lens focal length,  the total number of crystals N$_c$, arranged into rings or sectors, also  depends on the 
adjustable inter-distances between contiguous tiles, called tangential frame-width 
(w$_t$) and radial frame-width (w$_r$)\footnote[2]{The inter-distance between the 
tiles plays a crucial role in the definition of the lens $filling~factor$
that is the ratio between the area covered by the crystals and the total 
area covered by the lens cross section.}.

A specific library is devoted to the photon generation. The photons can be either uniformly distributed 
over the entire lens energy passband or distributed according to a given distribution curve (e.g., power-law).
Each generated photon is defined through 3 parameters: position $\vec{r_{i}}$,  wave vector $\vec{k}_i$ and energy $E_i$.
The total number of photons  N$_p$ is equally subdivided into the 
number of crystals, thus each crystal interacts with 
N$_p$ / N$_c$ photons. The generated photons are randomly distributed over the crystal surface. In the case of an astrophysical 
source, the assumption is a point--like source infinitely 
distant from the lens. In this case all the photons have the same wave vector $\vec{k}_i$. 

The code can also simulate an extended source at finite distance from the lens, which is the actual 
condition for the laboratory test of the lens. In this case, the  wave vector $\vec{k}_i$ depends on the 
coordinates of the position where the photon is generated within the extended source and on the coordinates 
of the position where the photon is incident on the crystal.

Once the lens geometry and the incoming photons are defined, the interaction is described by the Bragg law in 
vectorial form (central panel of Fig.~\ref{fig:diagram}) to get the propagation direction
of the emerging beam, and by the reflectivity formula to determine  the beam intensity as a 
function of both the diffraction angle and energy. 
The crystal reflectivity has been estimated by using  the dynamical theory of diffraction. For flat mosaic crystals we employed the treatise of Ref.~\citenum{Zachariasen}, for bent crystals we adopted have the theory reported in Ref.~\citenum{malgrange2002} 
which is an extension of the Penning and Polder theory~\cite{penning61}  of the X-ray diffraction
in crystals with curved diffracting planes. 

Concerning these crystals, the distortion of their diffracting planes due 
to the primary curvature  is described by the strain gradient:

\begin{align}
\beta = \frac{\Omega}{t ~ \delta/2}
\label{beta}
\end{align}

where $\delta = 2 d_{hkl}/\Lambda$ is the Darwin width of the crystal in which the extinction 
length $\Lambda$ is given by:

\begin{align}
\Lambda = \frac{\pi~ V ~cos~\theta}{r_e ~\lambda ~|C| ~|F_H|}
\label{lambda}
\end{align}
 
where $\lambda$ is the wavelength of the impinging radiation ($\lambda = hc/E$, where $h$ is the Planck constant and c is the light speed), $d_{hkl}$ is the 
spacing between the selected diffraction planes,  $V$ is the volume of the crystal unit cell, $r_e$ is the classical 
electron radius, $F_H$ is the structure factor, and $C$ is the polarization factor.
 
\begin{figure}[!h]
  \centering
    \includegraphics[scale=0.5]{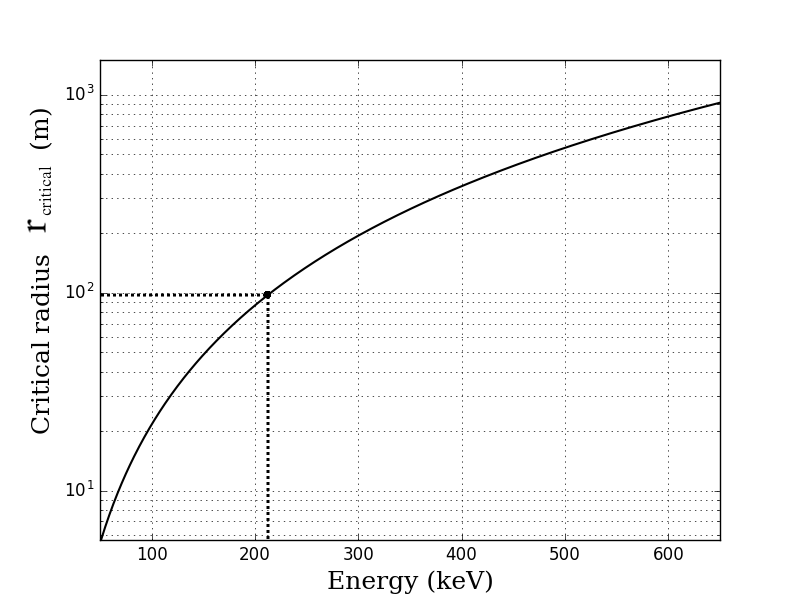}
  \caption{Critical primary curvature radius as a function of the energy. The highlighted point 
  represents the case of Ge (111) bent crystals with secondary curvature of 96.5~m that corresponds to a 
  critical energy of 211~keV.}
  \label{fig:E_rc}
\end{figure}

For a uniform curvature, when the strain gradient $\beta$ is larger than a critical value 
$\beta_c = \pi/2\Lambda$ which is inversely proportional to the energy, it has been shown~\cite{malgrange2002} that
the peak reflectivity, as a function of energy $E$ and secondary curvature, is given by:

\begin{align}
R^{peak}(r_s, E)  = \bigg( 1-e^{-\frac{\pi^2}{\beta~\Lambda}}  \bigg)~ e^{-\mu(E) ~t} = 
\bigg( 1-e^{-\frac{\pi^2 d_{hkl} r_s}{\Lambda^2}}  \bigg)~ e^{-\mu(E) ~t}
\label{Reflectivity} 
\end{align}

where $\mu(E)$ is the absorption coefficient.
For the validity of Eq.~\ref{Reflectivity}, it can be shown that the condition 
$\beta$ $>$ $\beta_c$ is equivalent to the following condition on the secondary curvature radius $r_s$:

\begin{align} 
r_s < r_{critical} = \frac{2 ~ \Lambda^2}{\pi~ d_{hkl}}
\label{r}
\end{align}

where also the critical radius $r_{critical}$ depends on photon energy, as shown in Fig.~\ref{fig:E_rc}, for a particular material and 
given diffracting planes. For instance, for a bent crystal made of Ge (111) with a primary 
curvature radius of 40~m (that results in a secondary curvature of r$_s$ $\sim$ 96~m), the critical energy E$_c$ = 211~keV, 
and Eq.~\ref{Reflectivity} can be applied down to this limit.  

\begin{figure}[!h]
  \centering
    \includegraphics[scale=0.5]{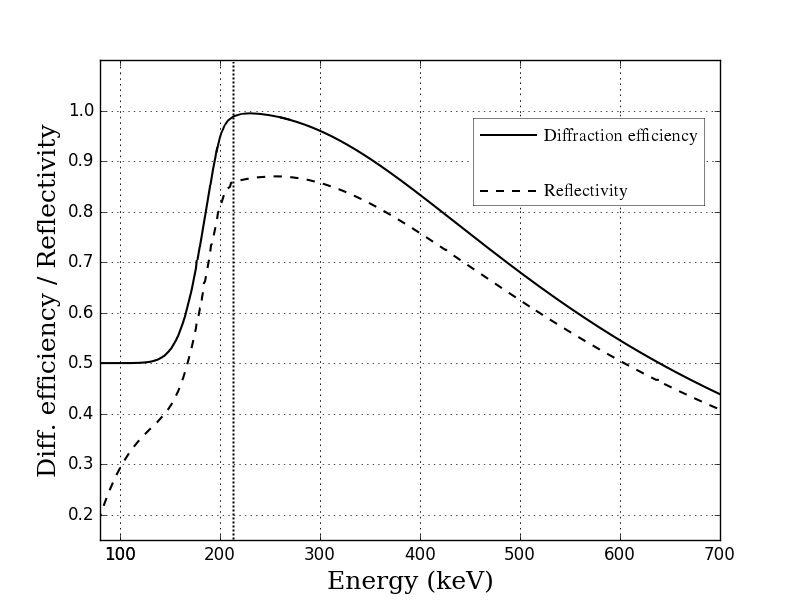}
  \caption{Simulated diffraction efficiency and reflectivity as function of energy in the 
  case of bent Ge (111) with curvature radius of 40~m.}
  \label{fig:reflectivity}
\end{figure}

The behavior of the reflectivity for $r_s>r_{critical}$ has been investigated by Ref.~\citenum{bellucci13b} using a multi-lamellar approach for the 
crystal, with the obvious condition that when $r_s \gg r_{critical}$ (i.e. flat crystal) the maximum diffraction efficiency is 0.5.  Experimental 
tests performed by these authors confirm the goodness of the followed approach. Adopting these results, in 
Fig.~\ref{fig:reflectivity} we show the expected diffraction efficiency and reflectivity in the case of Ge(111) with a primary curvature of 40~m.


Finally, following Ref.~\citenum{khalil15}, a position--sensitive detector located at a tunable distance $D$ from the lens is simulated 
in order to obtain the spatial distribution of the photons in the focal plane. 
Together with the spatial distribution of the photons, from the Monte Carlo simulations also the effective area and the sensitivity of the Laue lens are derived.

\section{The simulated lens}
\label{s4}

One of the main advantages of the Laue lenses is their flexibility in the design phase
thanks to the possibility of testing a large variety of crystals with different 
d-spacing that can be arranged at different ring radii of the lens. Nevertheless, 
the optimization of the lens design and the maximization of its effective area are tasks that 
lie outside of the goal of this paper, thus we have simulated a Laue lens capable of 
focusing photons in the 90--600 keV energy passband using a single crystal material.
The choice of a single material allows to study the radial distortion 
and the misalignment effects without to include other variables that directly or indirectly 
depend on the crystal material and that would make the discussion unclear.

\begin{table}[!t]
  \begin{center}  
  \caption{Parameters of the lens made by Ge(111) crystal tiles. }
  \vspace{0.2cm}
  \resizebox{11cm}{!}{
    \begin{tabular}{lll}
    \hline
 {\bf Lens properties}  & focal length	& 20~m 	\\
                        & energy pass band &  90 -- 600~keV	\\
                        & minimum lens radius	& 12.7~cm	\\
                        & maximum lens radius	& 93.7~cm 	\\
                        & number of rings	& 28		\\
                        & number of crystal tiles & 9341	\\
				  & filling factor		& 0.91  \\
                        \\
{\bf Crystal properties} ~~~~& material 	& Ge	\\
			& diffraction planes & (111)\\
                        & dimensions (l$\times$s$\times$t) 	& 30~mm $\times$ 10~mm $\times$ 2~mm \\	
                        \\             
			& crystal total mass        &    30~kg\\
    \hline
    \end{tabular}
    }
    \label{tab:Ge_lens_parameter}
  \end{center}
\end{table}

The main parameters of the simulated lens are described in Table~\ref{tab:Ge_lens_parameter}. The 20 m focal length Laue lens is made of 
Ge(111) crystal tiles with a size of $30\times 10 \times 2$~mm$^3$, arranged in 28 concentric rings from the innermost with radius 
r$_{in}\sim$~13 cm to the outermost with radius r$_{out}$ $\sim$ 94 cm. With these assumptions, and assuming a filling factor of 0.91,  the number of the lens crystal 
tiles is 9341 and the total crystal weight $\sim$ 30 kg. The crystal size is  the same adopted in our project \laue, so that 
a direct comparison between simulations and experimental results can be done. The 2~mm crystal thickness is dictated by the current 
technology which is not suitable for bending thicker crystals. The corresponding \qm\ spread of the crystal tiles, given 
by $\omega_m =t/r_s$, in our simulations is $\sim 4$~arcsec.

Assuming only paraxial photons, simulations have been performed by taking into account both crystal misalignments and 
radial distortion effects. Even if a Gaussian distribution would be more realistic,  
a uniform distribution was assumed, which gives an upper limit to their effects. 
Concerning the crystal misalignment, once defined a parameter $\delta$, we generated  for each crystal two random values of $\delta\alpha$ and $\delta\theta$ 
(see Sect.~\ref{effect_misalignment}) in the range ($-\delta$, $+\delta$).
For each lens simulation, a different value of $\delta$ was taken, between 
zero (each crystal properly oriented) and 30 arcsec (maximum misalignment). The value of 30 arcsec is a pessimistic value taking into account the technology we 
have developed for the realization of the Laue lens prototype.
Similarly, for the radial distortion, we introduced a parameter $\Delta$ and we generated for each crystal 
a random curvature radius in the range ($r - \Delta$, $r + \Delta$) where $r$
is the nominal curvature radius of the designed Laue lens.  
Progressively, the value of $\Delta$ was varied from zero (all the crystals have the  
nominal curvature radius $r$) to the maximum radial distortion of 6~m, which represents a pessimistic value of the primary 
curvature radius compared with the capability of the current developed technology.

\section{Simulation results}
\label{s5}

The effects of the simulated crystal misalignments and radial distortions were evaluated by deriving the 
half power diameter (\hpd) of the lens \psf, its FWHM, the peak reflectivity, the on--axis effective area and the corresponding sensitivity 
to continuum emission and to emission lines.

\subsection{Effect of crystal misalignment and radial distortion on the lens PSF}
\label{PSF}

\begin{figure}[!t]
\begin{center}
\includegraphics[scale=0.41]{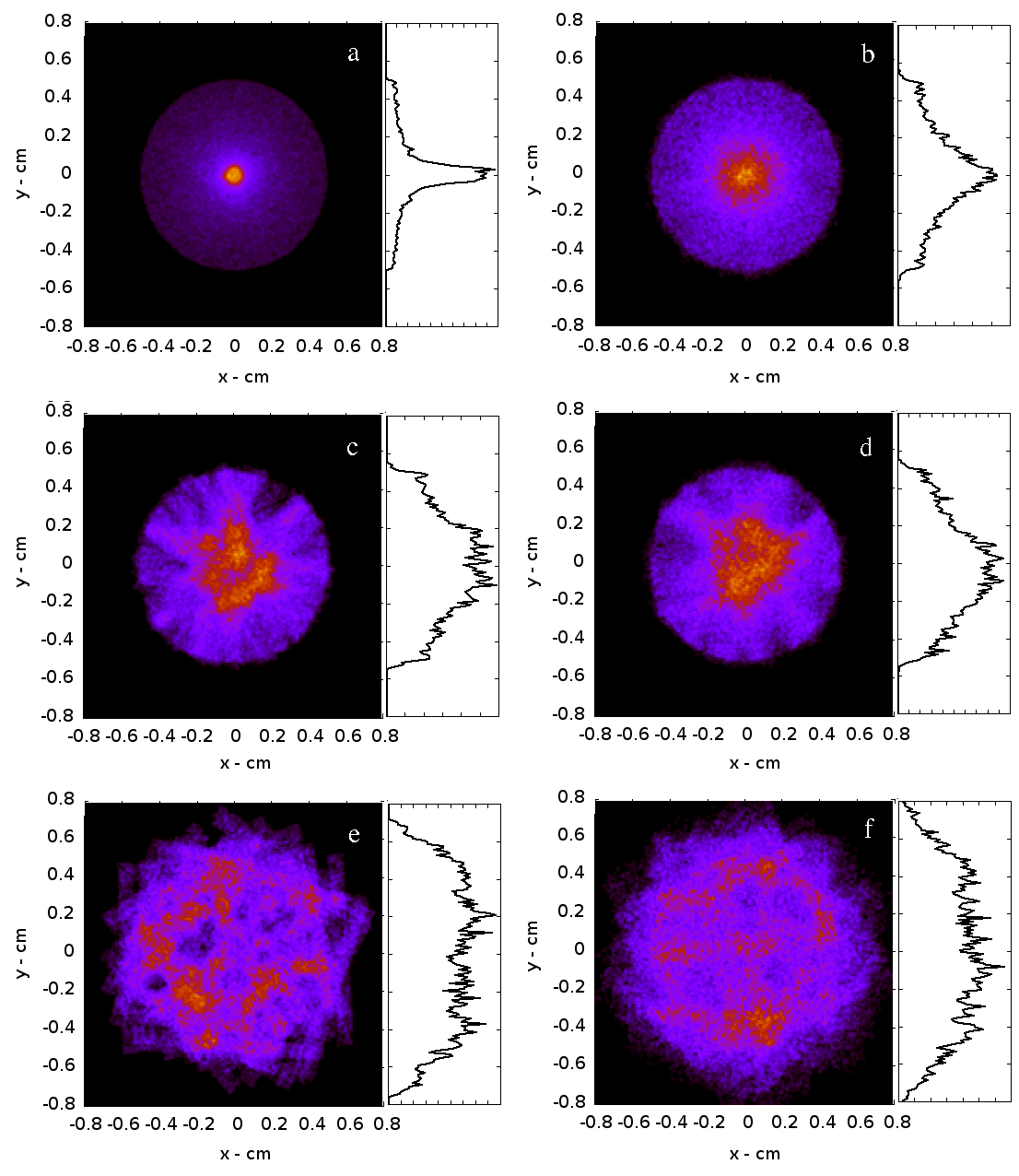}
\caption{\footnotesize{Expected photon distribution in the focal plane of a lens made with Ge(111): 
$\it a$) with no misalignment and no radial distortion; 
$\it b$) with a maximum radial distortion of $\pm 6$~m and no angular misalignment; 
$\it c$) with no radial distortion and 10~arcsec maximum crystal misalignment; 
$\it d$) with a maximum radial distortion of $\pm 6$~m and 10~arcsec maximum crystal misalignment;
$\it e$) with no radial distortion and 30~arcsec maximum crystal misalignment;
$\it f$) with a maximum radial distortion of $\pm 6$~m and 30~arcsec maximum misalignment of the crystals. 
The simulated images are assumed to be acquired with 
a 160 $\times$ 160 pixel PSD with pixel size of 100 $\times$ 100 $\mu$m$^2$.
The \psf\ profiles on the right of each plot are not to scale.}}
\label{psf_all}
\end{center}
\end{figure}

In Fig.~\ref{psf_all} we  show the images obtained from the Laue lens described in Table~\ref{tab:Ge_lens_parameter} in different cases: 
a) no radial distortion and no angular misalignment of the crystal tiles; 
b) a maximum radial distortion of $\pm 6$~m and no angular misalignment of the tiles; 
c) no radial distortion and a 10~arcsec maximum misalignment of the crystals; 
d) a maximum radial distortion of $\pm 6$~m and a 10~arcsec maximum misalignment of the crystals;
e) no radial distortion and a 30~arcsec maximum misalignment of the crystals; 
f) a maximum radial distortion of 
$\pm 6$~m  and a 30 arcsec maximum misalignment.

\begin{figure}[h!]
\centering
  \includegraphics[scale=0.55,keepaspectratio=true]{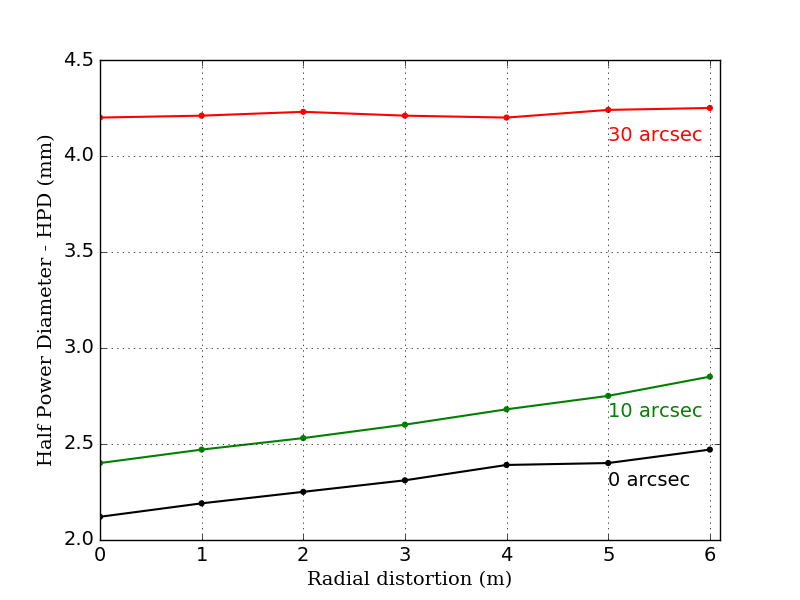}
  \caption{Values of the \hpd\ of the \psf\ for different values of the crystal 
  misalignment and radial distortion for a Laue lens made with Ge(111).}
  \label{fig:hpd}
\end{figure}

A broadening of the diffracted image is apparent when the radial distortion and/or the crystal tile misalignment are considered, with a larger effect ascribable to the tile misalignment  
with respect to the effect caused by the radial distortion. In Fig.~\ref{fig:hpd}
it is shown the \hpd\ of the lens \psf\ for 
different values of the maximum radial distortion and misalignments. The \hpd\ of the \psf\ 
increases with the crystal misalignment 
as well as with the radial distortion. When there is no misalignment and no radial distortion 
the \hpd\ is $\sim$ 2.1~mm which  
increases to 2.5~mm for the maximum radial distortion and no misalignment. If 
the misalignment effect is also introduced it critically affects the \hpd\ of the \psf, which increases
to $\sim$4.20~mm.

\begin{figure}[h!]
\centering
  \includegraphics[scale=0.55,keepaspectratio=true]{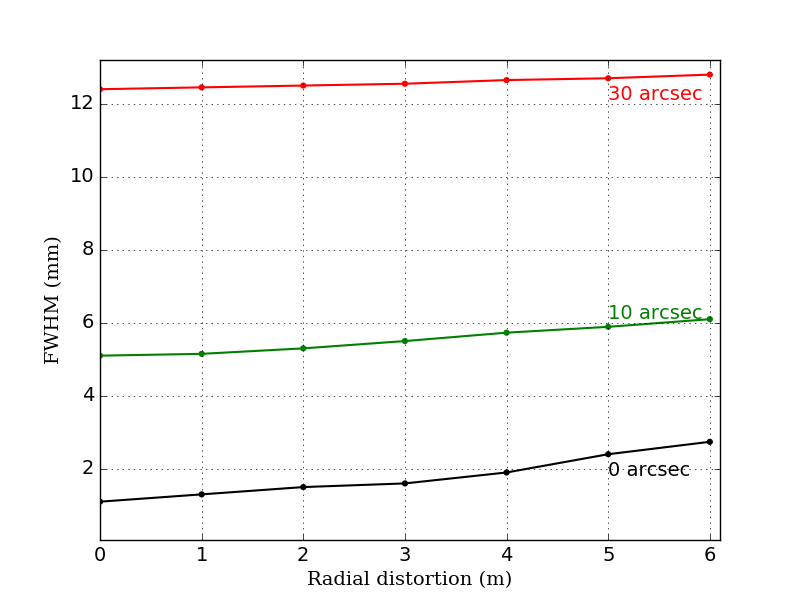}
  \caption{Values of the \fwhm\ of the \psf\ for different values of the crystal 
  misalignment and radial distortion for a Laue lens made with Ge(111).}
  \label{fig:fwhm}
\end{figure}

Also the \fwhm\ of the \psf\ has been determined.
The results are shown in Fig.~\ref{fig:fwhm}. When both the flaws are neglected, it results to be
\fwhm $\sim$ 1.1~mm,  which is consistent with the \fwhm\ obtained through Eq.~\ref{width2.39} 
for a 2 mm thick crystal.  It is worth noting that Eq.~\ref{width2.39} is valid for a single crystal but, in the case
of both null distortions and null angular misalignments,
the Gaussian profiles of the crystals superpose perfectly on top of each other 
resulting in a minimum increase of the combined \fwhm.

\subsection{Effect of the crystal misalignment and radial distortion on the normalized peak intensity}

\begin{figure}[!h]
  \centering
   \includegraphics[scale=0.55,keepaspectratio=true]{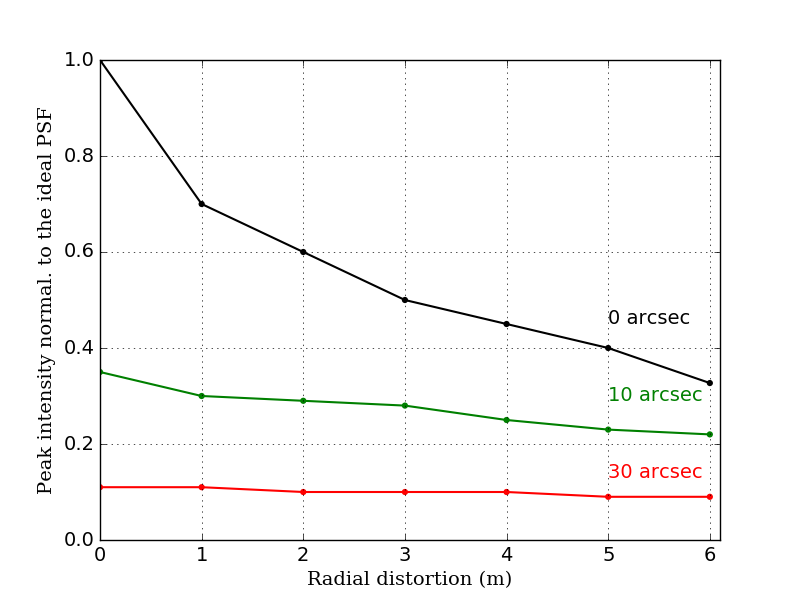}
  \caption{Peak intensity profile of the Laue lens made with Ge(111) for different 
  values of the crystals misalignment and radial distortion. The values in the legend 
  represent the maximum misalignment in the positioning of the crystals.}
  \label{fig:L_Ge_PI}
\end{figure}

Crystal misalignments and radial distortions also affect the peak intensity of the diffracted beam.
Figure~\ref{fig:L_Ge_PI} shows the normalized peak intensity  as a function of the crystal misalignment 
and radial distortion, including also the case of a nominal assembling
of the lens, with no misalignment and no radial distortion. 
As can be seen, in the case of no misalignment (black curve) the normalized peak intensity
rapidly degrades with the radial distortion until it reaches a half of its value when the radial distortion is about 3 m.
Independently of the value of the radial distortion, the peak intensity dramatically decreases to $\sim$10-15\% and remains almost flat in the plot
when the maximum misalignment is  30 arcsec, confirming that a high accuracy in the assembling of the crystal tiles
is crucial. 

\subsection{Effect of the crystal misalignment and radial distortion on the lens effective area}
\label{sec:EA}

The optimization of the lens effective area in a given energy passband is generally achieved
using different crystal materials with optimized 
thickness, depending on the energy to be diffracted (see, e.g., Ref.~\citenum{barriere09}).
However, given that the purpose of this paper
is to analyze how the effective area is sensitive 
to the variation of the radial and misalignment aberrations, we do not face the issue on how to maximize the lens effective area. 

\begin{figure}[!h]
\centering
\includegraphics[scale=0.43]{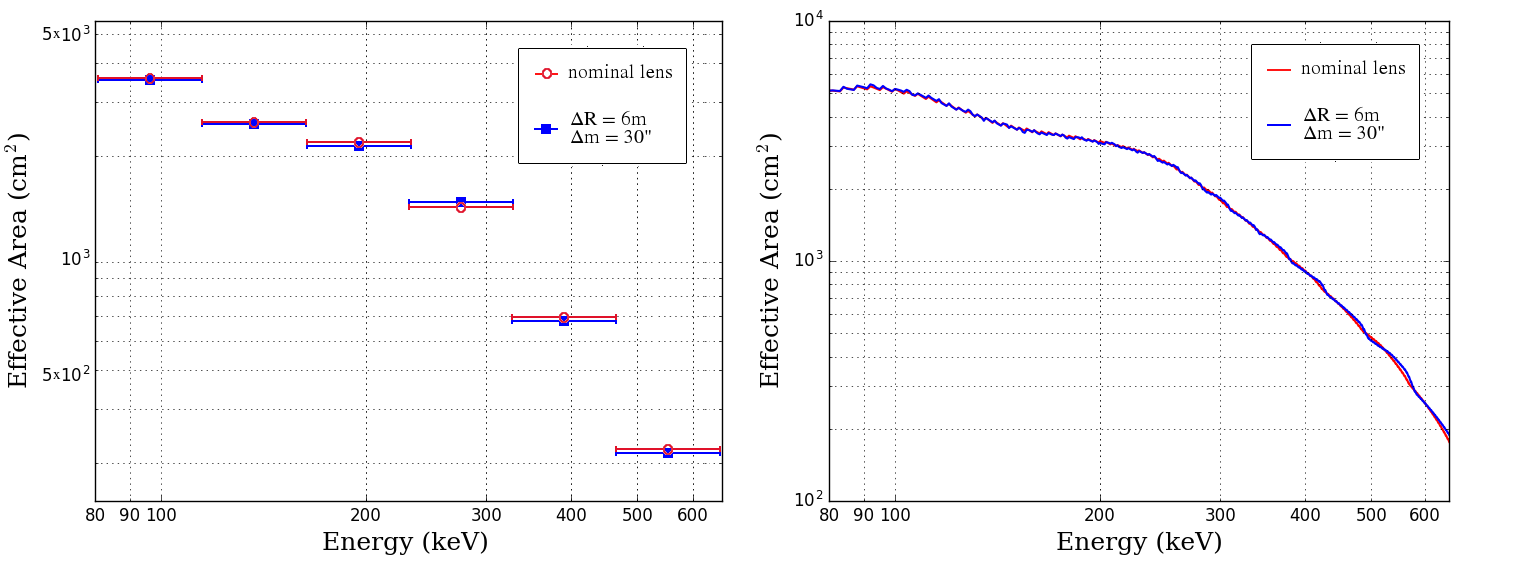}
  \caption{Expected effective area with energy of a Laue lens made with bent Ge (111) tiles, with overall 90-600 keV passband and focal length of 20 m 
(see Table~\ref{tab:Ge_lens_parameter}).
Left panel: 6 equal logarithmic energy bins. Right panel:
energy bins with $\Delta$E = E/2.
The effective area is shown both in the case of nominally bent and perfectly arranged tiles (red curves) and in the case of crystals radially distorted in the range $\pm$ 6~m with respect to the nominal value of 40~m and 
uniformly misaligned in the range $\pm$ 30~arcsec (blue curves).}
\label{fig:8}
\end{figure}

The effective area resulting from our simulations in the 90-600 keV is shown in Fig.~\ref{fig:8}, in 
the case of 6 energy bins of equal logarithmic width (left panel), and in the the case of bins of width 
$\Delta$E = E/2 (right panel). The value of the effective area changes considerably with the increase of 
the energy. This is mainly due to the number of crystal tiles that progressively decreases from outer to 
 inner radii. In fact, 
the number of Ge(111) crystal tiles settled in the ring corresponding to the lowest energy pass 
band (90 - 94 keV) is 502 while the inner ring 
devoted to the 537-681 keV  energy pass band contains only 70 crystals.
In both cases, the effective area is presented for both  a lens nominally 
built and for a lens affected by the maximum radial distortion 
and crystal tile misalignment. 
%

We point out that the effects of the two misalignment angles on the effective area are different.
While a rotation $\delta \alpha$ along the $x$ axis affects the photon distribution in the focal plane 
(as already pointed out in Sect.~\ref{PSF}), it does 
not affect the Bragg angle, thus it does not modify the diffracted energy. On the opposite side, $\delta \theta$ 
is the only responsible of the modification of the effective area.

\subsection{Expected detection efficiency and instrument background}
\label{sec:det_eff}

\begin{figure}[!t]
 \centering
 \includegraphics[scale=0.45]{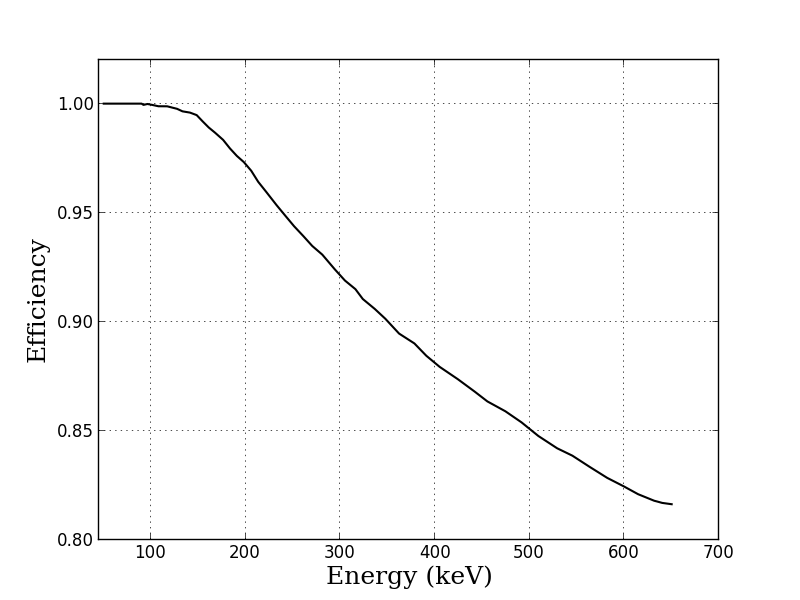}
 \caption{Detection efficiency as a function of the energy for the Germanium 
 detector assumed  in the simulations.}
 \label{fig:det_eff}
\end{figure}

The detection efficiency and the background level are crucial for the sensitivity estimate of a focusing instrument. In our code, both the detection efficiency and the detector background have been modelled to simulate a real detector.
Solid state Germanium detectors are good candidates  
as focal plane position sensitive detector (PSD)  for Laue lenses, as discussed elsewhere \cite{khalil15}.
On the basis of the properties of the simulated Laue lens, a detector with cross section 
of $10\times 10$~cm$^2$ and pixel size of 350 $\mu$m $\times$ 350 $\mu$m is a good solution \cite{khalil15}. 
A detection thickness of 12.5 cm or higher assures
a  high detection efficiency (80\%) up to the highest energies of the lens passband. 
The detector  efficiency $\eta_d$(E) adopted for the simulations is shown 
in Fig.~\ref{fig:det_eff}. 

\begin{figure}[t!]
 \centering
    \includegraphics[scale=0.45]{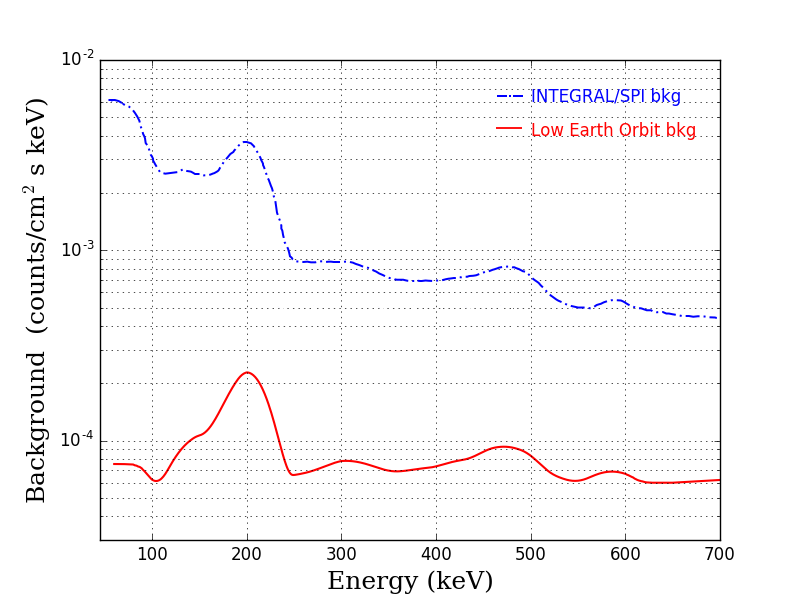}
 \caption{The expected lens background at LEO, compared with the INTEGRAL SPI measured background.}
 \label{fig:bkgd}
\end{figure}

The instrumental background for a Laue lens that is supposed to be operative in a Low Earth Orbit (LEO)
was estimated considering the background data in the 
90--600~keV energy band 
measured by SPI/INTEGRAL in a High Earth Orbit and extrapolating them 
to the LEO~\cite{dean89, khalil15} . The results are reported in Fig.~\ref{fig:bkgd} where 
the expected lens background is shown together with the SPI measured background, 
for comparison.

\subsection{Expected continuum sensitivity}
\label{sec:sens}
\begin{figure}[!h]
  \centering
 \includegraphics[scale=0.5]{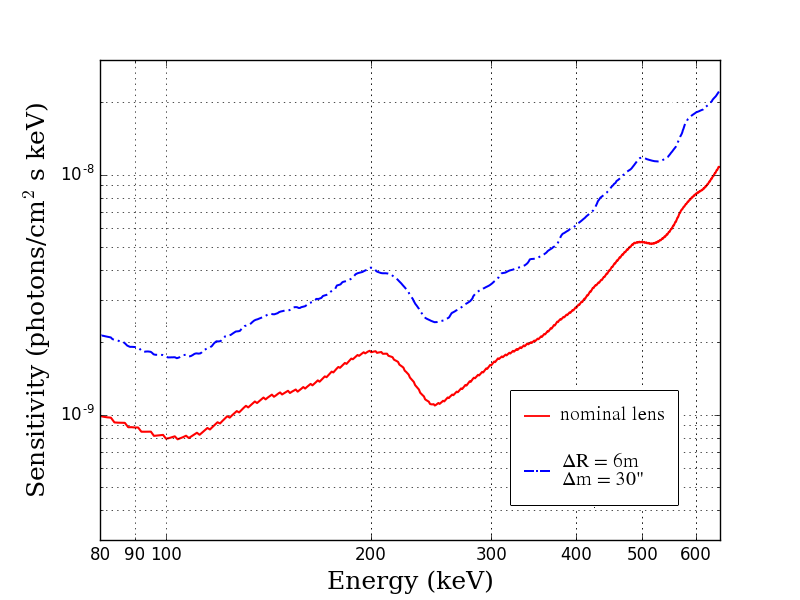}
 \caption{Expected on-axis continuum sensitivity ($3\sigma$ level) of the 
 simulated Laue lens made of Ge(111) bent crystal tiles in the 90-600 keV 
 energy pass band in the case of crystals with correct primary curvature radius and 
 without any misalignment (red curve) and in the case of a Laue lens 
 made of crystal tiles with maximum radial distortion of 
$\pm 6$~m and a maximum misalignment in the crystal positioning of 30 arcsec (blue curve). 
The simulation is made with T$_{obs}$ = 10$^5$~s and $\Delta E = E/2$.}
 \label{fig:SensitivityLaue}
\end{figure}

The minimum continuum intensity $I^{min}$ which is detectable by a focusing telescope at the confidence level corresponding to $n_\sigma$
is given by the expression:
 
\begin{equation}
   I^{min}(E) = n_{\sigma}~\frac{\sqrt{B(E)}~\sqrt{A_d}}{\eta_d(E)~f_{\epsilon}~A_{eff}~\sqrt{\Delta E}~\sqrt{T_{obs}}} 
   \label{eqn:CS_ft}
\end{equation}
 
where f$_{\epsilon}$ is the fraction of photons that is focused in the detector area A$_d$, 
T$_{obs}$ is the exposure time, A$_{eff}$ is the effective area in the considered
energy bin of width $\Delta$E around E, B(E) and $\eta_{d}$(E) are the assumed background 
intensity and the efficiency of the position sensitive detector, respectively. 
For our simulated Laue lens with a 90--600~keV passband, its continuum sensitivity at $3\sigma$ confidence level, 
with T$_{obs}$ = 10$^5$s, $\Delta$E = E/2, and $f_\epsilon$ = 0.5, which corresponds to $A_d = \pi (HPD/2)^2$, is plotted in 
Fig.~\ref{fig:SensitivityLaue}.  The red curve represents the case of perfect tile alignment with no radial 
distortion while the blue curve represents the case of tiles 
whose primary curvature radius  follow  a uniform distribution with maximum distortion 
of $\pm 6$~m and the crystals are uniformly misaligned with a maximum value of 30 arcsec.

As can be observed in Eq. \ref{eqn:CS_ft}, the worsening of the Laue lens sensitivity, when affected by the crystals inaccurate curvature and 
maximum inexact alignment, depends on both the effective area and its \psf\ evolution. Nevertheless, 
the range of the considered misalignments and distortions does not significantly affect the efficiency of the 
crystals therefore the overall intensity of the diffracted photons is unaltered. In fact, 
the effective area is essentially the product of the crystals efficiency with the lens geometric 
 area, thus it  does not suffer a significant alteration, 
 as confirmed by the simulations presented in Sect.~\ref{sec:EA}. It is worth noting that, 
 in order for the above consideration to be valid, the detector size must be 
 large enough to contain all the diffracted photons. On the contrary, the angular distribution of the reflected photons 
 is dramatically affected by the considered shortcomings, i.e. the 
 \psf\ size plays, in these range of the considered shortcomings, a dominant role on the Laue lens sensitivity.  

\subsection{Expected sensitivity to narrow emission lines}

For a focusing telescope, the sensitivity to a narrow emission line, in photons/(cm$^2$~s), 
superposed  to the continuum source spectrum  at a confidence level corresponding to $n_\sigma$, is given by:

\begin{equation}
  I_L^{min}(E_l) = 1.31 n_{\sigma}\frac{\sqrt{[2 B(E_l) A_d + I_c(E_l) \eta_{d} f_{\epsilon} A_{eff}]\Delta E} } { \eta_{d} f_{\epsilon} A_{eff} \sqrt{T_{obs}}} 
 \label{eqn:LS_ft}
\end{equation}

where $E_L$ is the line centroid,$I_c(E_L)$ is the source continuum intensity at the line centroid, $\Delta$E is the \fwhm\ of the line profile and depends upon the energy resolution of the detector which has been 
assumed to be 2 keV (expected for our simulated detector \cite{khalil15}). The other parameters, all calculated at the energy $E_L$, are those defined in the previous section.

At a confidence level corresponding to $3\sigma$, for both cases of a Laue lens unaffected by shortcomings and a Laue lens with uniform distribution of the radial 
distortion (max $\pm 6$~m) and tile misalignment (max 30 arcsec), the expected line intensity results is shown in Table~\ref{table2} at different energies and 
for an observation time of $10^5$~s. For comparison, 
the line sensitivity achievable with the IBIS and SPI 
instruments~\cite{Roques03} on board INTEGRAL is also reported. As it can be seen, in spite that the simulated lens is not optimized as discussed above, the improvement factor of the simulated Laue lens with respect to INTEGRAL is very large at low energies and is still significant at high energies.

\begin{figure*}[!t]
  \centering
  \includegraphics[scale=0.4]{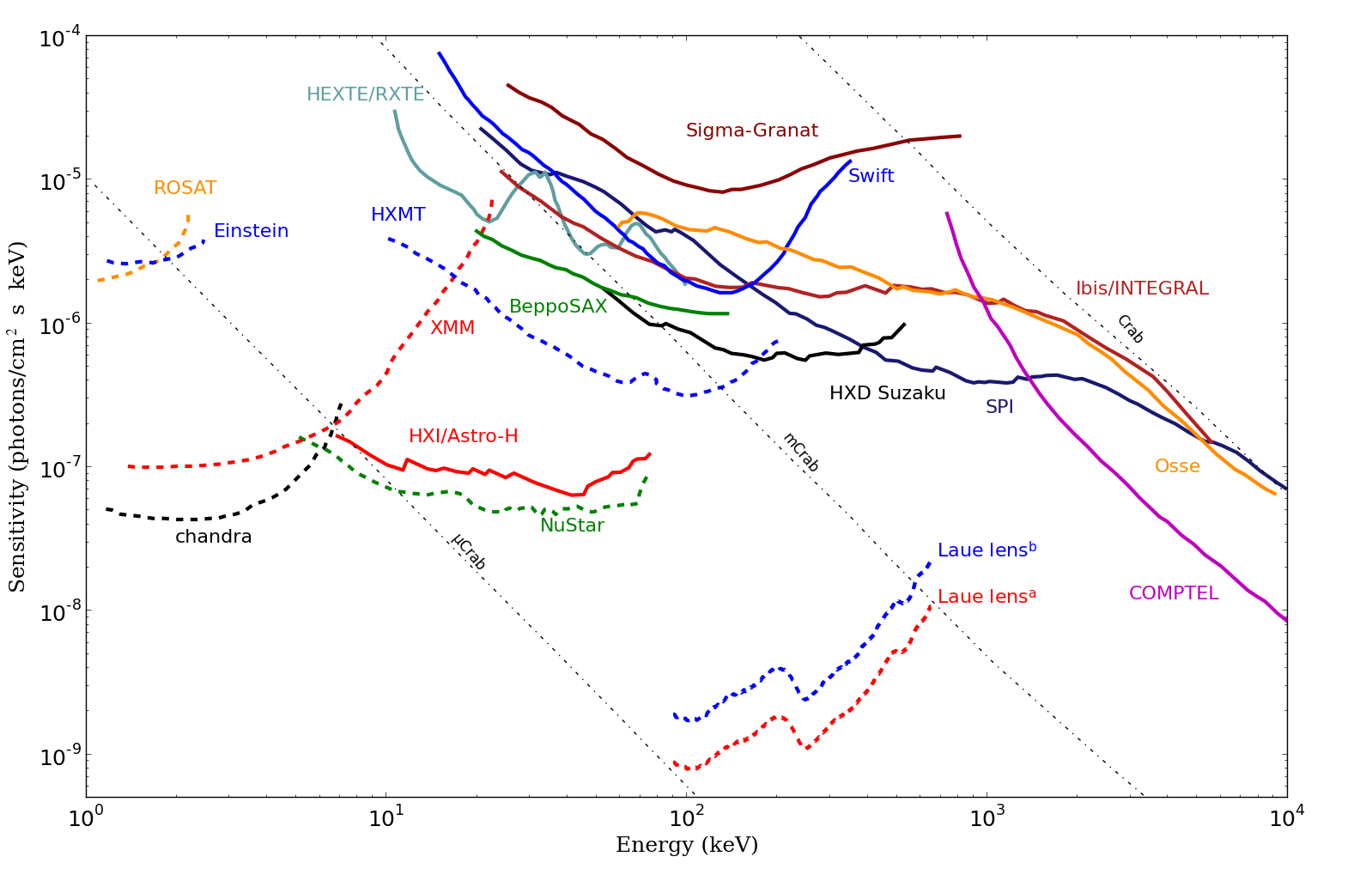}
 \caption{The $3\sigma$ sensitivity with T$_{obs}$ = 10$^5$~s and $\Delta E = E/2$ of past and present direct-view instruments (continuous lines),and 
of the past and current focusing telescopes (dashed lines) in the X--/Gamma--ray domain, 
 compared with the sensitivity achievable with the simulated Laue lens. The lens sensitivity is given in the two cases shown in Fig.~\ref{fig:SensitivityLaue}.} 
 \label{fig:SensitivityMissions}
\end{figure*}

\begin{table}[!h]\renewcommand{\arraystretch}{1.3}\addtolength{\tabcolsep}{0pt}
\begin{center}
\caption{\footnotesize $3 \sigma$ line sensitivity (in photons/s/cm$^2$) of a Laue lens in the 90-600 keV energy pass band with an observation time of $10^5$~s for both the case 
of an ideally assembled Laue lens and a Laue lens suffering from a maximum radial distortion
of 6 m and a maximum misalignment of 30 arcsec. For comparison, the sensitivity of INTEGRAL/ISGRI 
derived from~\cite{Roques03} and of INTEGRAL/SPI (derived from the Ibis Observer's Manual) are also reported.} 
\label{table2}
\vspace{0.4cm}
\resizebox{13cm}{!}{
\medskip     
\begin{tabular}{cccccc} 
 \hline
    Energy  &      simulated lens      &   simulated lens        &   INTEGRAL   &    INTEGRAL       & improvement factor \\
    (keV)  &         $^{(a)}$     &    $^{(b)}$        &   ISGRI      &      SPI          &   \footnotesize{SPI / lens$^{(a)}$}\\
    \hline
100     &  3.7$\times$ 10$^{-7}$& 7.4 $\times$ 10$^{-7}$& 6.0 $\times$ 10$^{-5}$ &  1.3 $\times$ 10$^{-4}$   &   351 \\
200     &  2.3$\times$ 10$^{-6}$& 4.6 $\times$ 10$^{-6}$& 1.0 $\times$ 10$^{-4}$ &  2.6 $\times$ 10$^{-4}$   &   113\\
300     & 4.6 $\times$ 10$^{-6}$& 9.3 $\times$ 10$^{-6}$& 1.4 $\times$ 10$^{-4}$ &  1.1 $\times$ 10$^{-4}$   &   23\\
400     & 1.2 $\times$ 10$^{-5}$& 2.3 $\times$ 10$^{-5}$& 1.8 $\times$ 10$^{-4}$ &  1.1 $\times$ 10$^{-4}$   &    9\\
500     & 2.3 $\times$ 10$^{-5}$& 4.9 $\times$ 10$^{-5}$& 2.5 $\times$ 10$^{-4}$ &  1.4 $\times$ 10$^{-4}$   &    6\\   
\hline
\end{tabular}
}
\end{center}
\footnotesize{
$^{(a)}$ Perfectly aligned crystal each with the nominal curvature radius.\\
$^{(b)}$ Crystals affected by a maximum radial distortion of 6~m and by a maximum misalignment of 30~arcsec.}
\end{table}

\section{Conclusions} 
\label{s6}

We have modelled and simulated a Laue lens made of bent Ge(111) crystal tiles  with 20~m focal length and an 
energy passband from 90 to 600 keV. The on-axis \psf, its Half Power Diameter, \fwhm, 
effective area and sensitivity have been determined in two cases: a perfect lens with no crystal radial 
distortion and no crystal tile misalignment, and a realistic lens in which crystal tile misalignments 
and distorted crystals are present.

In Fig.~\ref{fig:SensitivityMissions} it is shown the expected continuum sensitivity of the simulated 
lens, compared with that of past and still operational X--/Gamma--ray instruments (focusing or not).
In spite of a still not optimized configuration, the expected sensitivity of the simulated lens is 
very high in the 90--600 keV passband.
This is due to the large effective area of 
the lens along with the high focusing effect enabled by bent crystals, even in the case of radial 
distortion and a non perfect alignment of the crystals.  

Also the expected sensitivity to emission lines of the simulated lens has been investigated. As shown 
in Tab.~\ref{table2}, up to 200 keV a large improvement (more than two orders of magnitude) with respect 
to the INTEGRAL IBIS and SPI instrument sensitivity is found. This  sensitivity  is very 
important, e.g., for a deep study of the 158 keV Ni$^{56}$ line emitted at the early epoch of 
Type 1a supernovae (see, e.g., Ref.~\citenum{Isern14}). However an optimization of the 
lens at higher energies is recommended for  deep nuclear science studies, like that 
of the origin of the positron annihilation line at 511 keV from the Galactic 
Center region  (see, e.g., Ref.~\citenum{Prantzos11}).  This optimization can be 
achieved with a larger diameter of the lens and a selection of the best 
crystal thickness and material as in the case of the ASTENA instrument concept. 
With ASTENA we expect to propose in the next few years 
an advanced satellite mission that we are confident it will lead to an unprecedented 
leap forward in the study of the hard X--/soft gamma--ray sky.

\acknowledgments
We wish to thank the anonymous referees for their very useful suggestions and comments. 
The authors acknowledge the Italian Space Agency for its support 
to the \laue\ project under contract I/068/09/0. VV acknowledges the 
support from Erasmus Mundus Joint Doctorate Program by Grant Number 2010-1816 from  
the EACEA of the European Commission. Currently, VV is now supported by the Indian TMT Project.

\bibliographystyle{spiejour}
\bibliography{virgilli2017}

\end{document}